\documentclass[amssymb,amsfonts,aps,pra,superscriptaddress,twocolumn,showpacs]{revtex4-1}

\usepackage{graphicx}
\usepackage{dcolumn}
\usepackage{bm}

\begin{document}


\title[Sample title]{Applicability of scaling behavior and power laws in the analysis of the magnetocaloric effect in second-order phase transition materials}

\newcommand{\uamftmc}{Departamento de F\'{i}sica Te\'{o}rica de la Materia Condensada, Universidad Aut\'{o}noma de Madrid, E-28049 Madrid, Spain}
\newcommand{\victorino}{Departamento de F\'{i}sica de la Materia Condensada, ICMSE-CSIC, Universidad de Sevilla P.O. Box, 1065, 41080 Sevilla, Spain}
\newcommand{\tanaka}{Waseda Institute for Advanced Study, Waseda University, 1-6-1 Nishi-Waseda, Shinjuku-ku, Tokyo 169-8050, Japan}
\newcommand{\tamura}{Computational Materials Science Unit, National Institute for Materials Science (NIMS), 1-1 Namiki, Tsukuba, Ibaraki 305-0044, Japan}

\author{Carlos Romero-Mu\~{n}iz}
\email{carlos.romero@uam.es}
\affiliation{\uamftmc}

\author{Ryo Tamura}
\affiliation{\tamura}

\author{Shu Tanaka}
\affiliation{\tanaka}

\author{Victorino Franco}
\affiliation{\victorino}

\date{\today}

\begin{abstract}
In recent years, universal scaling has gained renewed attention in the study of magnetocaloric materials. It has been applied
to a wide variety of pure elements and compounds, ranging from rare earth-based materials to transition metal alloys; from bulk
crystalline samples to nanoparticles. It is therefore necessary to quantify the limits within which the scaling laws would
remain applicable for magnetocaloric research. For this purpose, a threefold approach has been followed: a) the magnetocaloric
responses of a set of materials with Curie temperatures ranging from 46 to 336 K have been modeled with a mean-field Brillouin model, b) experimental data for Gd has been analyzed, and c) a 3D-Ising model ---which is beyond the mean-field approximation--- has been studied. In this way we can demonstrate that the conclusions extracted in this work are model-independent. It is found that universal scaling remains applicable up to applied fields which provide a magnetic energy to the system up to 8\% of the thermal energy at the Curie temperature. In this range, the predicted deviations from scaling laws remain below the experimental error margin of carefully performed experiments. Therefore, for materials whose Curie temperature is close to room temperature, scaling laws at the Curie temperature would be applicable for the magnetic field range available at conventional magnetism laboratories ($\sim10$ T), well above the fields which are usually available for magnetocaloric devices.

\end{abstract}


\maketitle

\section{Introduction}

The magnetocaloric effect (MCE) is an intrinsic property of some materials consisting in a reversible temperature change upon the application/removal of a magnetic field. Nowadays, it represents a hot topic in the material science research. This is mainly due to its applications in refrigeration technologies which are increasingly becoming more competitive \cite{kitanovski2014}. The main advantages of the magnetic refrigeration are related to its environmental benefits compared to older established technologies. It is thought, especially after the discovery of several materials with a giant magnetothermal response \cite{pecharsky1997,wada2001,fujieda2002}, that the viability of devices based on MCE materials is feasible, with a variety of prototypes which have been developed in recent years \cite{yu2010}. During this period of thorough research in this field, many materials with remarkably MCE have been found. They belong to different families: intermetallic compounds, Heusler alloys, amorphous materials, ceramic manganites, etc. In parallel, the physical foundations of this phenomenon have been clarified both from the theoretical and experimental points of view as shown in several review papers and books \cite{franco2012,smith2012,oliveira2010,tishin2003}.

Because of the practical importance of the MCE there has been a remarkable increment in the number of papers devoted to the modeling of this effect from a theoretical point of view \cite{shen2009,oliveira2010,franco2012}. These theoretical studies on the MCE were carried out in order to estimate the main MCE parameters such as magnetic entropy change and adiabatic temperature change. It is worth noting that these models allow us to extract useful information about the foundations of the MCE and they can provide detailed and complementary information to the experimental results, especially in the high applied field regions, where experiments are normally more difficult to perform. A wide variety of models exist in the theoretical study of the MCE. Among these theoretical models we can distinguish between two main groups.

On the one hand we find the so called first principles methods which try to obtain the magnetothermal response without any empirical information, or at least with the minimum amount of fitting parameters. These methods consist on purely microscopic approaches. Two classical methods belonging to this group are quantum mechanics Hamiltonians solved by Monte Carlo (MC) algorithms and density functional theory (DFT) simulations. Both of them have already been used in modeling MCE. There are some interesting papers in which MC simulations are used with different materials like Gd(Si$_x$Ge$_{1-x}$)$_4$ \cite{nobrega2005}, (Gd$_x$Tb$_{1-x}$)$_5$Si$_4$ \cite{nobrega2006}, and several Heusler alloys \cite{buchelnikov2010,singh2013,comtesse2014}. Recently, a new type protocol of the MCE was proposed in theoretical studies based on the MC methods \cite{tamura2014a,tamura2014b}. On the other side, DFT is less used in this research field, although it has been used to describe both the structural and the magnetic phase transitions in some rare-earth intermetallic pseudo-alloys exhibiting a giant MCE \cite{paudyal2006,paudyal2007,singh2009,singh2010,paudyal2011,mudryk2012}, and there have been very few attempts to describe the MCE of materials from first principles by only using DFT. At the moment, this latter approach only succeeded with simple metals like gadolinium \cite{staunton2014}. Note that both of the mentioned methods consist in microscopic approaches to obtain the MCE.

The other group of models is derived from phenomenological laws like those of molecular field or the Landau theory. In this kind of approaches, which are mainly from a macroscopic point of view, the final result is an equation of state in which magnetization $(M)$, applied field $(H),$ and temperature $(T)$ are related by an equation of the form $\Omega(M,H,T)=0.$ Although this expression is analytical it is not ensured that it can be solved for each variable separately. Of course, it is always possible to treat the cited function numerically to obtain values of magnetization for different temperatures and applied fields. In this way, the MCE properties can be easily estimated in the same way used for experimental data [i.e. by numerically processing the $M(H,T)$ data]. These phenomenological equations of state are divided in two main groups. The first group includes those equations derived from mean-field approximations, like the molecular field theory by P. Weiss \cite{weiss1907}, the generalization proposed by C. P. Bean and D. S. Rodbell \cite{bean1962} or the equation more recently derived by M. D. Kuz'min \cite{kuzmin2008} based on the Landau theory. The second group is constituted by those expressions derived from the scaling relations including critical exponents like the widely used Arrott-Noakes equation \cite{arrott1967} and others less used like the one derived by J. T. Ho and J. D. Litster \cite{ho1969} a couple of years later. Furthermore, it is possible to find some examples of generalizations of  mean-field equations of state to include other critical exponents, as it was proposed again by A. Arrott a few years ago \cite{arrott2010}.

In any case, all these models must fulfill some scaling laws in the proximity of the critical point. These scaling laws were deduced by B. Widom in his, now famous, paper of 1965 \cite{widom1965}, where he showed how the variation of some properties followed a certain scaling law governed by some powers called critical exponents. Although there are many exponents involving different magnitudes, only two of them are independent. All others can be obtained through the so called scaling relations. During the following years, the ideas of B. Widom were experimentally confirmed in many systems including magnetic materials and fluids \cite{heller1967,vicentinimissoni1970}. Of course, all this scaling formalism can be transferred to the study on the MCE, since the maximum in the magnetic entropy change occurs normally near the critical point. For this reason, it is possible to find power-law dependencies of the main MCE parameters with respect to the applied field in materials with a second-order phase transition \cite{franco2010}. This scaling behavior has been observed experimentally many times in a wide range of materials including rare earths, transition metals based alloys, amorphous alloys, and manganites \cite{pekala2010,dhahri2012,debnath2013,wang2013,su2013,sharma2013}. Recently, it was even used to predict the concentration of skyrmions \cite{ge2015}.

However, this behavior is supposed to be present only in the vicinity of the critical point. During the last years, the magnetocaloric community has been using these scaling laws to fit and predict the values of some MCE properties regardless of the conditions of work temperatures and applied fields. The purpose of this work is to clarify under which conditions it is possible to use these scaling laws. For completeness, in Sec. \ref{theory}, we will start with an overview of the theory and methodologies which will be followed by their applications to different kind of materials and models. In Sec. \ref{results}, we will analyze the results obtained by using a mean-field approximation in a wide selection of ferromagnetic materials with different Curie temperatures. This allows us to derive the range of applicability of the scaling laws in the MCE in terms of temperature and applied field. In order to demonstrate its validity, we will include some experimental data to support our hypothesis. Then, we will carry out a similar analysis using a 3D-Ising model, which is a microscopic treatment beyond the mean-field approximation. We will extract similar conclusions, proven that are independent of the model used. Finally, we will make some brief remarks about other models used in MCE research regarding how they behave with respect the critical scaling.

\section{Theory}\label{theory}

\subsection{Scaling behavior and field dependence}\label{scaling}

Before starting with the analysis of the scaling behavior and the field dependence of some MCE quantities, we must remember that the MCE is characterized by the magnetic entropy change and the adiabatic temperature change that a sample undergoes upon the application of a magnetic field. The first quantity can be defined through the integrated version of the Maxwell relation as:
\begin{equation}
\Delta S_{\rm{M}}(T,H) = \mu_0 \int_0^{H} \left(\frac{\partial M}{\partial T} \right)_{H'} dH'.
\end{equation}
And the second one is defined as:
\begin{equation}
\Delta T_{\rm{ad}}(T,H) = -\mu_0 \int_0^{H} \frac{T}{c(T)_{H'}}\left(\frac{\partial M}{\partial T} \right)_{H'} dH',
\end{equation}
where $c(T)_H$ is the specific heat at constant field of the material and $\mu_0$ is the permeability of vacuum.

In this work we will only focus on the magnetic entropy change. Probably the first attempt to describe the field dependence of the maximum magnetic entropy change was made over thirty years ago by H. Oesterreicher and F. T. Parker \cite{oesterreicher1984} who showed that $-\Delta S_{\rm{M}}(T_{\rm{C}},H) \propto H^{2/3}$ by expanding the Brillouin function on a power series, where $T_{\rm{C}}$ is the Curie temperature. This empirical law was widely accepted among the magnetocaloric community. Later, M. D. Kuz'min \cite{kuzmin2009} added a small negative term independent of $H$, which arose from spatial inhomogeneities of real ferromagnetic materials. J. Lyubiuna et al. \cite{lyubina2011}, based on the Landau theory, proposed a more complex dependence (including the Kuz'min constant term) as:
\begin{eqnarray}
-\Delta S_{\rm{M}}(T\approx T_{\rm{C}},H) &=& A(H+H_0)^{2/3} \nonumber \\
&& -AH_0^{2/3} + BH^{4/3},
\end{eqnarray}
where $A$ and $B$ are intrinsic material constants and $H_0$ is related with the Kuz'min constant term. Similar results were obtained by using other mean-field approximations like the Green functions formalism applied by P. \'{A}lvarez et al. \cite{alvarez2011}. In an equivalent way, similar relations were derived for the adiabatic temperature change \cite{kuzmin2011}. Independently from these mean-field approximations, in 2006, V. Franco et al. \cite{franco2006} proposed that the field dependence of the magnetic entropy change near $T_{\rm{C}}$ was a power law with an exponent $n$ related to the critical exponents of the material: $n=1+(\beta-1)/(\beta+\gamma).$ The $n = 2/3$ case is a particular case of this general expression when we choose the mean-field critical exponents $\beta=1/2$ and $\gamma=1.$ It was proved experimentally that materials with critical exponents far from the mean-field approximation obeyed this power law \cite{franco2008b,mnefgui2014}. The exponent $n$ can be calculated for all temperatures from the magnetic entropy change curves and not only at the critical point, using the following expression
\begin{equation}\label{exponent}
n(T,H) = \frac{d\ln|\Delta S_{\rm{M}}(T,H)|}{d\ln H}.
\end{equation}

For a given applied field, the behavior of this exponent is as follows. It has a minimum near $T_{\rm{C}}$, (exactly at $T_{\rm{C}}$ for mean-field approximation) whose value is the one pointed out previously depending on the critical exponents ($2/3$ in the frame of the mean-field approximation). For temperatures well above the Curie temperature, in the paramagnetic region, it reaches the value of 2. On the contrary, for temperatures below the Curie temperature, it reaches the value of 1 in the purely ferromagnetic region \cite{franco2010}.

It is also possible to construct a unique normalized magnetic entropy change curve for all values of applied field, which was initially called master curve \cite{franco2006} and eventually universal curve \cite{Franco2008}. The collapse of all magnetic entropy change curves for different applied fields has been proved that it is a consequence of the critical scaling behavior \cite{Franco2008}. For magnetic systems near the critical point, a scaling relation between magnetization, applied field, and temperature must be fulfilled and it has the form \cite{griffiths1967}:
\begin{equation}\label{grif}
\frac{H}{M^{\delta}} = f(tM^{-1/\beta}),
\end{equation}
where $\delta$ and $\beta$ are the critical exponents, $t\equiv (T-T_{\rm{C}})/T_{\rm{C}}$ and $f$ is a scaling function which depends on the model or the material. For mean-field models, this scaling is fulfilled with $\delta=3$ and $\beta=1/2.$ Notice that this relation should be fulfilled only in a region close to the phase transition and not in the whole range of applied field and temperature, except for some models which has been constructed exclusively from scaling hypothesis. As expected, the universal curve has been applied to different materials exhibiting a second-order phase transition. However, materials with first-order phase transitions do not collapse onto a universal curve and this fact can be used to determine the nature of the phase transition in a given material \cite{bonilla2010}. In order to construct this phenomenological curve, we firstly normalize the magnetic entropy change dividing by the maximum $\Delta S_{\rm{M}}/\Delta S_{\rm{M}}^{\rm{pk}}.$ Then we choose two reference temperatures, which must fulfill the following conditions: $\Delta S_{\rm{M}}(T_{\rm{r1}}<T_{\rm{C}})/\Delta S_{\rm{M}}^{\rm{pk}} = \Delta S_{\rm{M}}(T_{\rm{r2}}>T_{\rm{C}})/\Delta S_{\rm{M}}^{\rm{pk}} = h,$ where $0<h<1$ is an arbitrary constant. Although in principle $h$ could be freely selected between 0 and 1, a too large value (reference temperatures chosen too close to the peak temperature) would produce large numerical errors due to the limited number of points -- in experimental measurements -- which lie in that region. Conversely, if $h$ is too small, it implies selecting reference temperatures far from the critical region, where other phenomena could take place. Once the two reference temperatures are found, we define a new variable $\theta$ for the temperature axis as:
\begin{equation} \label{universal}
\theta = \left\{ \begin{array}{ll} -(T-T_{\rm{C}})/(T_{\rm{r1}}-T_{\rm{C}}) & \textrm{if } T\le T_{\rm{C}} \\ \\
(T-T_{\rm{C}})/(T_{\rm{r2}}-T_{\rm{C}}) & \textrm{if } T> T_{\rm{C}}. \end{array} \right .
\end{equation}

The representation of the different magnetic entropy change curves on the $\Delta S_{\rm{M}} /\Delta S_{\rm{M}}^{\rm{pk}}$ and $\theta$ axis produces the phenomenological universal curve. It was subsequently proved that the use of two reference temperatures was not necessary, unless there were multiple phases in the sample or the demagnetizing factor was not negligible \cite{franco2010}. Therefore, in our analysis, we will only use a single reference temperature for $T>T_{\rm{C}}.$ Notice that in Eq. (\ref{universal}), we have used $T_{\rm{C}}$ which coincides with $T^{\rm{pk}}$ in mean-field models but in general they might be slightly different \cite{franco2009}. In any case, they are very close and they can be used indistinctly.

\subsection{Mean-field model}\label{meanfieldtheory}

One simple approach to describe the ferromagnetic behavior of a substance is to consider the possible orientations of the magnetic moment with respect the applied field $H$ at temperature $T,$ according to the possible $2J+1$ values of the magnetic quantum number (from $-J$ to $+J$). In that case, it is straightforward to calculate the single-dipole partition function \cite{pathria1996}:
\begin{equation}
Z(T,H,M) = \frac{\displaystyle{\sinh\left(\frac{2J+1}{2J}x\right)}}{\displaystyle{\sinh\left(\frac{1}{2J}x\right)}},
\end{equation}
where $x=g\mu_0\mu_{\rm{B}}J(H+\lambda M)/k_{\rm{B}}T.$ Here $k_{\rm{B}}$ is the Boltzmann constant, $g$ is the Land\'{e} factor, $\mu_{\rm{B}}$ is the Bohr magneton and $\lambda$ is the phenomenological constant of the Weiss molecular field. From this partition function it is possible to deduce all important quantities, especially the magnetization:
\begin{equation}\label{brillouin}
M = \frac{Nk_{\rm{B}}T}{\mu_0}\left(\frac{\partial \ln Z}{\partial H}\right) = M_{\rm{s}}\mathcal{B}_J(x),
\end{equation}
where $M_{\rm{s}} = (N/V)\mu_{\rm{B}}gJ$ is the saturation magnetization and $\mathcal{B}_J(x)$ is the well know Brillouin function. Here, $N$ and $V$ are the number of magnetic ions and the volume, respectively. Notice that this expression represents an equation of state for the magnetic system and this model will be referred as mean-field Brillouin model. Although the scaling relation is not obtained for the whole range of temperature, it is not hard to deduce that for $T\approx T_{\rm{C}}$ the scaling relation given by Eq. (\ref{grif}) is indeed confirmed with $f(x) = x + c$ being $c$ a constant. The magnetic entropy per mole, in terms of the gas constant $R$, has the following analytical expression:
\begin{eqnarray}
\frac{S_{\rm{M}}(T,H,M)}{R} &=& \ln Z + T\left(\frac{\partial \ln Z}{\partial T}\right)  \nonumber \\ \nonumber \\  \nonumber
&=&
\ln\frac{\displaystyle{\sinh\left(\frac{2J+1}{2J}x\right)}}{\displaystyle{\sinh\left(\frac{1}{2J}x\right)}}
-x\mathcal{B}_J(x). \\
\end{eqnarray}

The magnetic entropy change can be calculated as $\Delta S_{\rm{M}} = S_{\rm{M}}(T,H) - S_{\rm{M}}(T,0).$ At very high temperatures the magnetic entropy reaches its maximum value (maximum disorder) and it is easy to see that $S_{\rm{M}}(T\rightarrow\infty,H) = R\ln(2J+1).$ By applying a magnetic field all dipoles begin to align in the same direction of the applied field, hence reducing the magnetic entropy, reaching  $S_{\rm{M}}(T\rightarrow0,H) = S_{\rm{M}}(T,H\rightarrow\infty) = 0$ for very low temperature or infinitely high field. Therefore, the maximum magnetic entropy change achievable for a given material is $-R\ln(2J+1)$. In this work we do not use the analytical form of the magnetic entropy to calculate the MCE. Instead of this, we proceed in an analogous way as that used for experimental data; differentiating and integrating the magnetization curves. In this way we get much more information about the system. This simple model only needs two parameters to describe the MCE; the saturation magnetization (or only the atomic density) and the Curie temperature. Remember that in this model, the Curie temperature is related to the phenomenological constant, as $T_{\rm{C}} = \mu_0(N/V)J(J+1)(g\mu_{\rm{B}})^2\lambda/3k_{\rm{B}}$. To study the field dependence of the magnetic entropy change peak we can use the scaling relation near the Curie temperature and it is possible to obtain the following expression:
\begin{equation}\label{fielddep}
\Delta S_{\rm{M}}(T_{\rm{C}},H) \simeq -\frac{\mu_0\lambda}{2T_{\rm{C}}}\left[\frac{10(J+1)^2M_{\rm{s}}^2}{3\lambda(2J^2+2J+1)}\right]^{2/3}\times H^{2/3}.
\vspace{0.3cm}
\end{equation}
As it was expected, the dependence is a power law with $n = 2/3,$ which corresponds with the mean-field critical exponents. This result was previously obtained by J. H. Belo et al. \cite{belo2012} and of course is compatible with the mentioned behavior suggested by H. Oesterreicher and F. T. Parker in 1984 \cite{oesterreicher1984}.

Despite being very simple, this mean-field model can be used to fit experimental data with a reasonable agreement in simple ferromagnetic materials \cite{oliveira2010}. This model can be improved by adding a new term in the argument of the Brillouin function of the type $\lambda_3M^3$ as proposed by C. P. Bean and D. S. Rodbell \cite{bean1962}. They showed this new term affected the transition temperature and, moreover, depending on the value of the phenomenological constant $\lambda_3,$ the transition could be of first or second-order type. On the other hand another extra term can be added to the Gibbs free energy to take into account changes of volume during the phase transition. In this way it is possible to study pressure effects during the phase transition which sometimes have an important role in the MCE. This model has been successfully used to reproduce the MCE of some materials with  giant MCE and first-order phase transitions where pressure may have an important role \cite{vonranke2004b,vonranke2004c,vonranke2006,rocco2007,alho2012}, and some other materials like manganites \cite{amaral2007} with both first and second-order phase transitions. Moreover this kind of theoretical models can provide useful information about the foundations of the different types of phase transitions that produce the MCE \cite{vonranke2005,oliveira2008}.

\begin{table*}[t]
\caption{\label{table1}%
Parameters used in the mean-field Brillouin model simulations for gadolinium based compounds. From crystallographic data it is possible to determine the number of atoms and the volume of the unit cell $V_{\rm{cell}},$ to calculate the mass density $\rho,$ and the saturation magnetization $M_{\rm{s}} = (N/V)\mu_{\rm{B}}gJ.$ }
\begin{ruledtabular}
\begin{tabular}{lcccccccccc}
Material & Structure & Space Group & $V_{\rm{cell}}$  & $\rho$   & $T_{\rm{C}}$  &   $\lambda$  & $g$  & $J$  & $M_{\rm{s}}$ at $\approx0$ K  & Refs. \\
         &           &             &   (\AA$^3$) & (g cm$^{-3}$) &      (K) &    &     &  & (kA m$^{-1}$) &       \\
\colrule
Gd$_5$Si$_4$  &	 Sm$_5$Ge$_4$  &  $Pnma$        &  854.78	&  6.98	    &  336    & 87.7 &  2	&  7/2  &	1513	&  \cite{altounian2007}\\
Gd	          &  hcp		   &  $P6_3/mmc$    &  66.101	&  7.90	    &  293    & 58.8 &	 2	&  7/2  &	1963	& \cite{legvold1980}\\
Gd$_3$In	  &  Fe$_3$C	   &  $Pnma$        &  391.561	&  9.95	    &  213    & 42.3 &	 2	&  7/2  &	1989	& \cite{hutchens1974}\cite{zhang2013}\\
GdMg          &	 ClCs		   &  $Pm\bar{3}m$  &  55.35	&  5.45	    &  119    & 40.1 &	 2	&  7/2  &	1172	& \cite{aleonard1975}\\
GdRu$_2$	  &  MgZn$_2$	   &  $P6_3/mmc$    &  214.241	&  11.14    &  84	  & 27.4 &  2	&  7/2  &	1212	& \cite{compton1959}\cite{andoh1987}\\
GdPd	      &  CrB		   &  $Cmcm$        &  183.419	&  4.77	    &  46	  & 25.7 &  2	&  7/2  &	707 	& \cite{hara2008}\\
\end{tabular}
\end{ruledtabular}
\end{table*}

\begin{figure*}[t]
\includegraphics[width=1.0\textwidth]{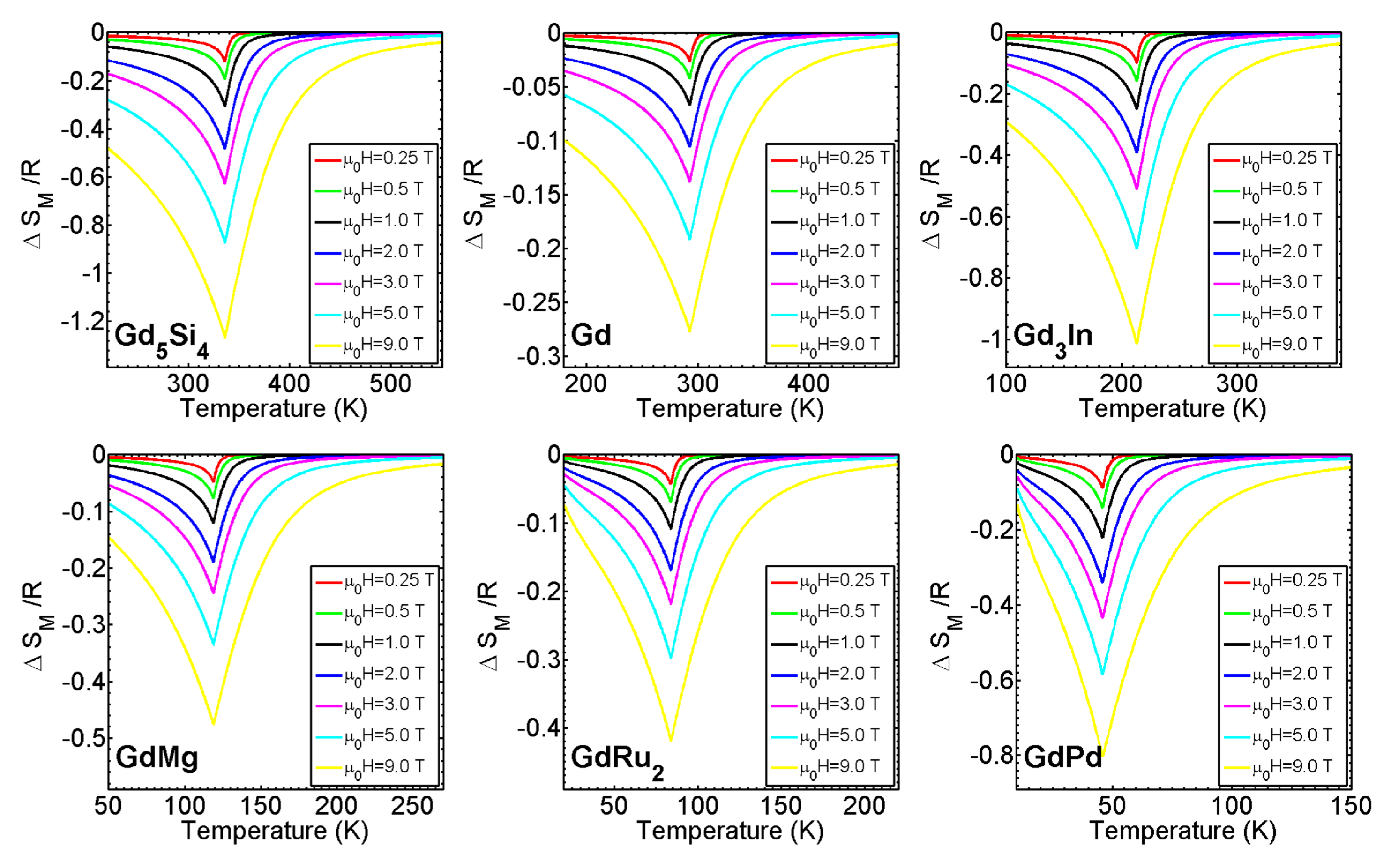}
\caption{\label{fig1} Magnetic entropy change obtained in the mean-field Brillouin model for the selected materials at several values of applied field. They are ordered by decreasing Curie temperature. Be aware that the scale of ordinate-axis is different depending on the material.}
\end{figure*}

\subsection{3D-Ising Model}\label{isingtheory}

The other model that we will use in this work is the Ising model on a cubic lattice which will be referred as 3D-Ising model. Let $N$ be the number of sites which is given by $N=L\times L \times L,$ where $L$ is the linear dimension. The Hamiltonian of this model is defined as:
\begin{equation}
\mathcal{H} = - J_{\rm{ex}}\sum_{\langle i,j \rangle} s_i^zs_j^z-g\mu_0\mu_{\rm{B}}H\sum_{i}s_i^z, \qquad (s_i^z = \pm 1/2),
\vspace{0.3cm}
\end{equation}
where $J_{\rm{ex}}>0$ is the ferromagnetic exchange interaction parameter between nearest-neighbor spin pairs $\langle i,j \rangle.$ Here we consider an $S=1/2$ system which corresponds to the case of $J=1/2$ in Sec. \ref{meanfieldtheory}. Thus, only at zero applied field, this model exhibits a second-order phase transition. According to Ref. \cite{ferrenberg1991}, in which high-precision MC calculation was done, the phase transition temperature, that is, the Curie temperature, is $k_{\rm{B}}T_{\rm{C}}/J_{\rm{ex}}=1.127$ and the critical exponents are $\beta = 0.3265,$ $\gamma = 1.2372,$ and $\delta = 4.789$ \cite{pelissetto2002,hasenbusch2010}. In this paper, we calculate the magnetic entropy change of this model by MC simulations based on the Wang-Landau method \cite{wang2001a,wang2001b,lee2006}. The Wang-Landau method is one of the multicanonical methods, which can directly obtain the density of states. Then, the behavior of magnetic entropy can be calculated with high accuracy \cite{tamura2014a,tamura2014b}. We confirmed that the magnetic entropies for $L=8$, $L=12,$ and $L=16$ are almost collapsed. Thus, we use the simulation results for $L=16$ in the analysis throughout this paper. Notice that the maximum value of magnetic entropy achievable for this model is $R\ln2 \approx 0.693R$ when $k_{\rm{B}}T \rightarrow \infty.$

\section{Results and discussion}\label{results}

\subsection{Mean-field model}\label{meanfieldresults}

\begin{figure*}[t]
\includegraphics[width=1.0\textwidth]{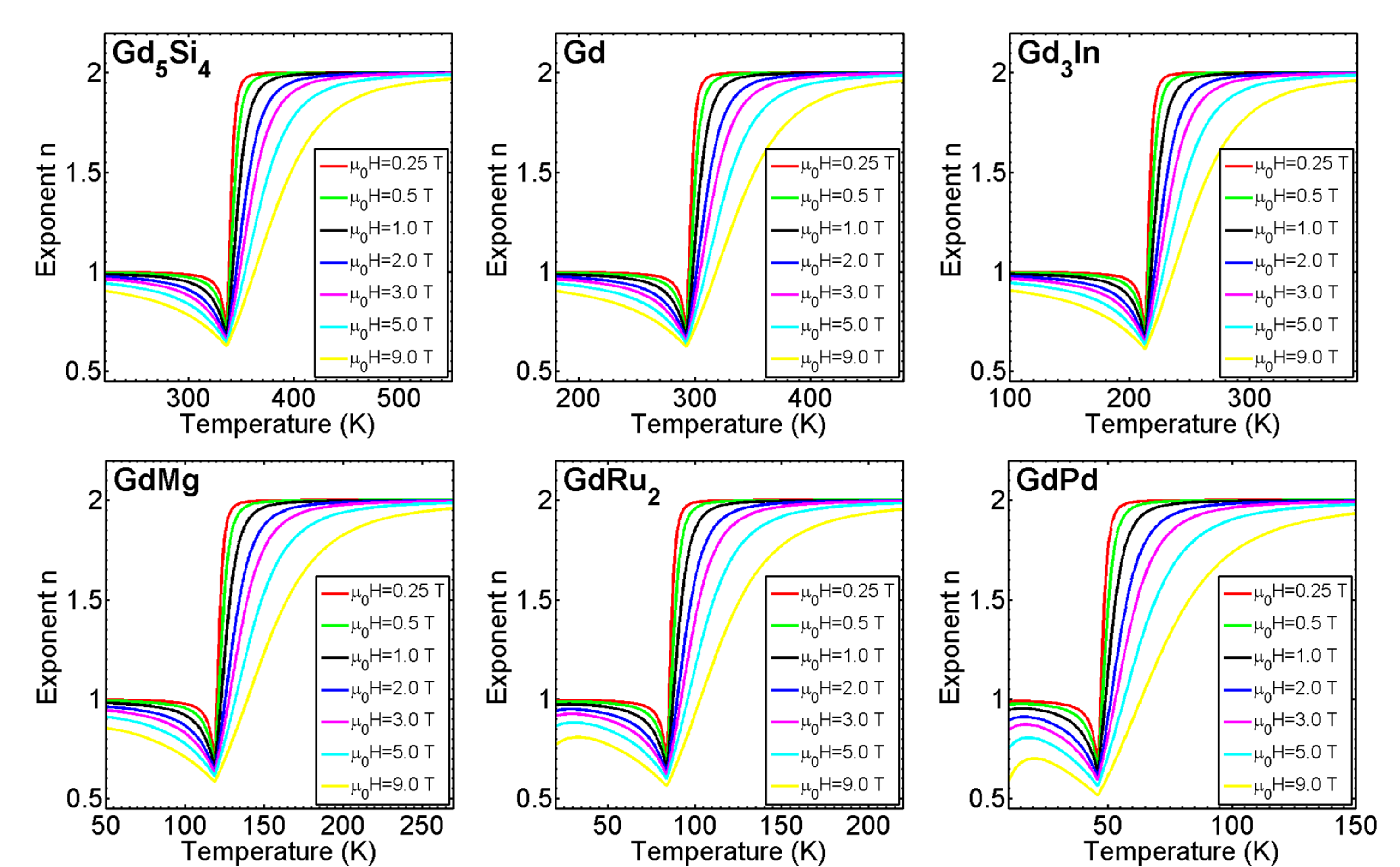}
\caption{\label{fig2} Exponent $n$ obtained in the mean-field Brillouin model through Eq. (\ref{exponent}) for the selected materials at different applied fields. Notice that for those materials with low Curie temperatures, that is, GdRu$_2$ and GdPd, the value of the exponent at $T_{\rm{C}}$ becomes field dependent for moderately high fields deviating from the mean-field predicted value of 2/3. Besides, for these low $T_{\rm{C}}$ materials at very low temperatures, the limit of $n = 1$ is not reached.
}
\end{figure*}

\begin{figure*}[t]
\includegraphics[width=1.0\textwidth]{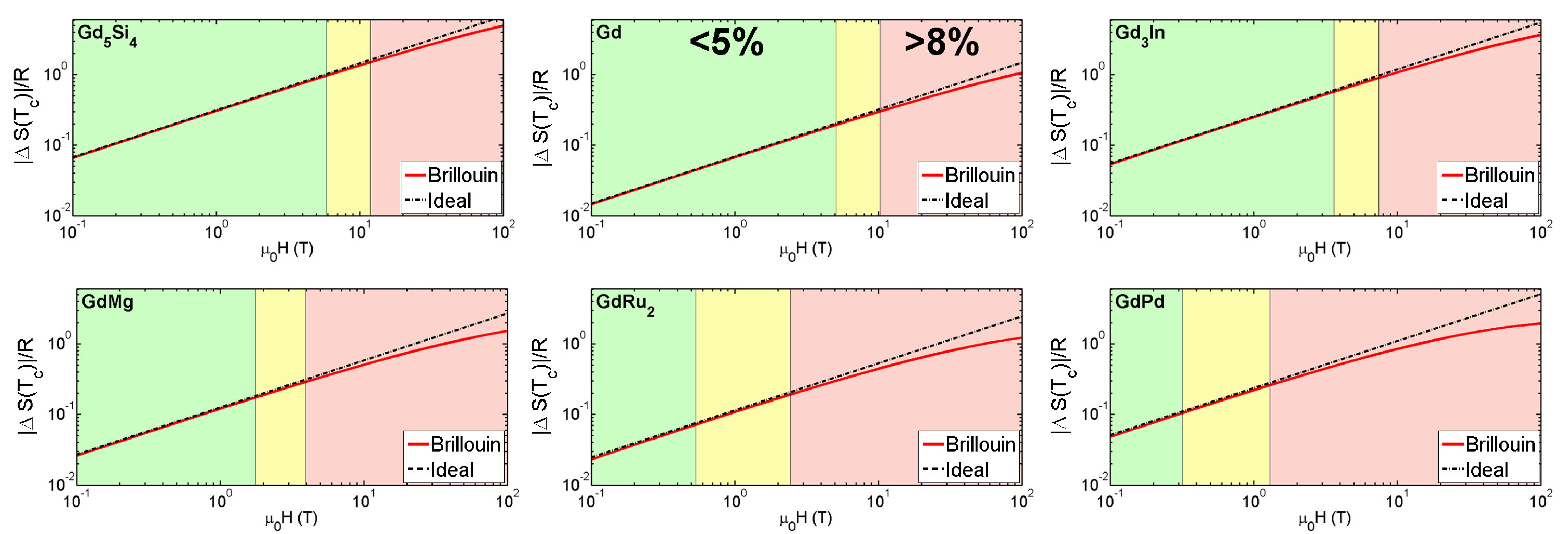}
\caption{\label{fig3} Applied-field dependence of the magnetic entropy change peak for the selected materials ordered by decreasing Curie temperature. In each graph, we have a comparison between the analytical expression of the ideal scaling (black dashed line) and the real behavior obtained with the mean-field Brillouin model (red solid line). In order to clarify the main results, all graphs are divided in three regions: low-field regime where the difference between the peak value predicted by the ideal relation and the real one is below a 5\% (green) so the scaling approximation is valid; high-field regime, where the scaling approximation is not valid any more because the discrepancies are over 8\% (red) and a mid-field region where the errors are within 5\% and 8\% (yellow).
}
\end{figure*}

In this subsection we will analyze the results regarding the mean-field Brillouin model for ferromagnetic materials. As it was previously pointed out, this model reproduces in a reasonable way the magnetic behavior of simple ferromagnetic materials. Therefore, in principle, we cannot deal with materials exhibiting other types of magnetic order such as antiferromagnetism, ferrimagnetism, etc. Among these ideal ferromagnetic materials we can find pure elements like iron, cobalt, and nickel. All lanthanides except lanthanum and lutetium have unpaired $f$ electrons so they do present magnetic behavior. All of them have been previously studied in this frame of the mean-field approximation \cite{tishin1990,tishin1998}. However, speaking strictly, the use of this simple model is only justified in the case of gadolinium, which has a simple ferromagnetic behavior. In all other cases, the crystal field interaction plays a key role in the establishment of the magnetic order. The mean-field model would be valid only in the high applied-field regime well above the critical fields of the antiferromagnetic order. Of course, there are more sophisticated formalisms available to deal with these materials \cite{oliveira2010}. Since we are restricted to a narrow group of materials we will focus on gadolinium and some gadolinium based compounds with non-magnetic atoms. We present in Table \ref{table1} our selection of materials covering a wide range of Curie temperatures from 46 K in GdPd to 336 K in Gd$_5$Si$_4$. In this way we are able to study how the magnitude of applied fields affects the scaling relations of the MCE in these materials with very different working temperatures. As it was shown in Sec. \ref{meanfieldtheory}, for this model, we only need the Curie temperature of the material and its saturation magnetization (or the atomic density, which can be easily calculated from crystallographic data). All required parameters are collected from the literature and compiled in Table \ref{table1}. It is worth mentioning that, in order to be able to make realistic predictions which would match experimental results, it is necessary to make such a broad selection of materials and transition temperatures. The reliability of the predictions will be shown with experimental results for Gd in Sec. \ref{experimentsresults}.

\begin{figure*}[t]
\includegraphics[width=1.0\textwidth]{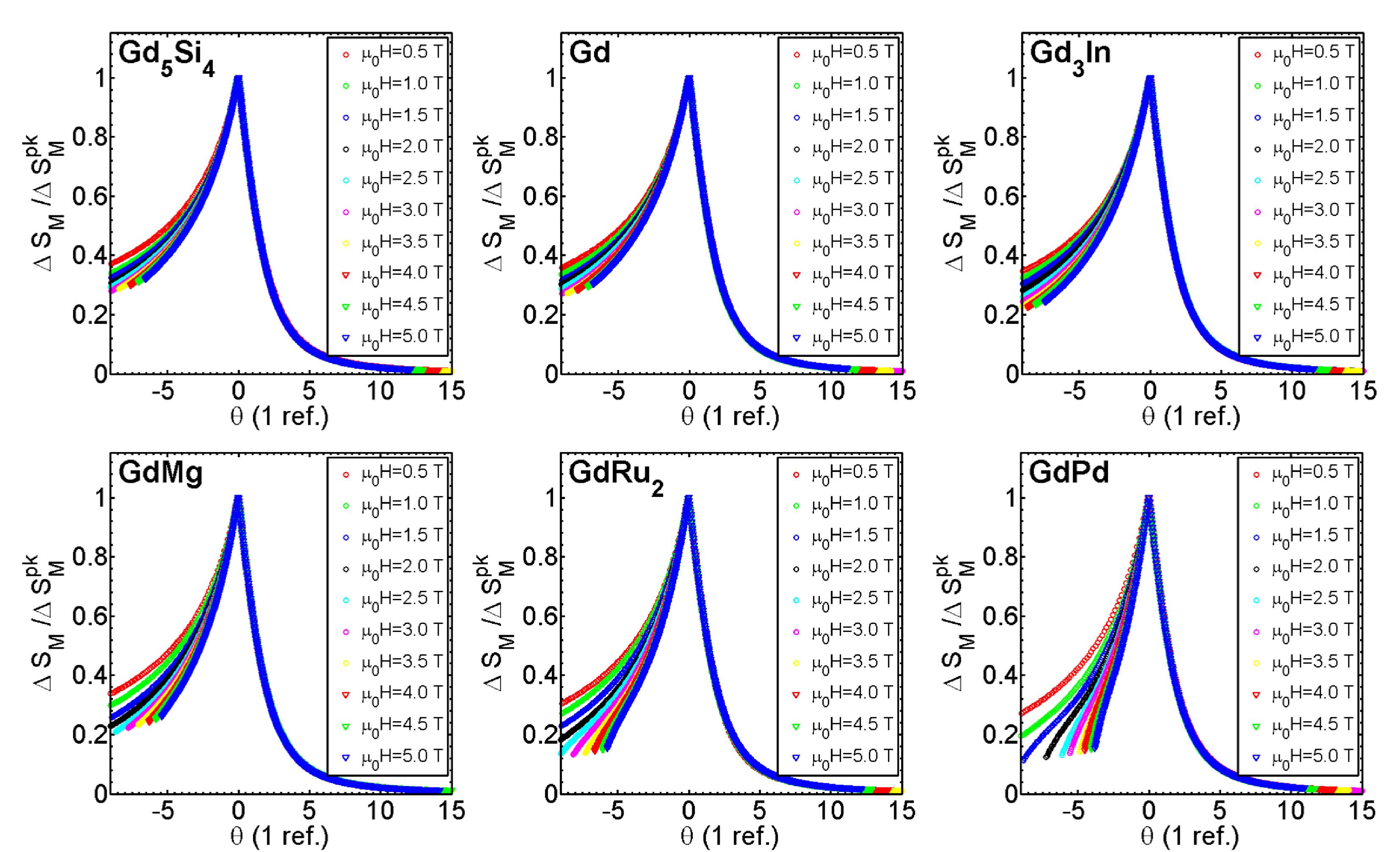}
\caption{\label{fig4} Normalized magnetic entropy change curves as a function of the reduced temperature $\theta$ using a single reference temperature defined with $h=0.6$ for the selected materials. Notice that the larger deviations from scaling are observed for the materials with low $T_{\rm{C}}.$
}
\end{figure*}

The calculated magnetic entropy change curves in the mean-field Brillouin model for different applied fields are shown in Fig \ref{fig1}. Since the magnetic entropy is expressed per mole, the magnitude of the MCE is quite different depending on the material, ranging from $-0.3R$ to $-1.2R$ for 9 T, still far from $R\ln8 \approx 2.08R$ of the maximum value. All of the curves have the well know caret shape characteristic of simple ferromagnetic materials with a second-order phase transition. Just after this preliminary characterization, we can calculate the exponent $n$ for the whole range of temperatures, as shown in Fig. \ref{fig2}. Although it seems that all materials follow the behavior described in Sec. \ref{scaling}, if we analyze these results carefully we can appreciate how some deviations from the expected behavior appear in those materials with low Curie temperatures. For instance, the ideal tendency of $n = 2$  for $T\gg T_{\rm{C}}$ is found in all cases, but this limit is reached at higher temperatures for increasing values of applied field. However, if we focus around the Curie temperature, we can see a slight decrease of $n$ at $T_{\rm{C}}$ for increasing fields. In principle, since we are using a mean-field approximation, $n$ was expected to be 2/3 in all cases. This is true for Gd$_5$Si$_4$, Gd, Gd$_3$In, and GdMg (with minor discrepancies). But for GdRu$_2$ and GdPd, whose Curie temperatures are 84 K and 46 K respectively, the drop in the value is quite evident, being close to $n=0.5$ for 9 T in both cases. Moreover, if we look at the ferromagnetic region, (below $T_{\rm{C}}$) for these low Curie temperature materials the limit of $n=1$ is not reached. In fact, at some point in temperature $n$ starts to decrease. After this brief analysis, we could wonder which are the applicability limits of the power laws in the study of the MCE. Of course, it is well know that all the power laws described in Sec. \ref{scaling} are only fulfilled in the ``neighborhood of the phase transition.'' Our goal is to transform this loose statement into a more quantitative one when the MCE magnitudes are concerned.

\begin{table}[t]
\caption{\label{table2}%
Relevant values of applied field for the applicability of scaling relations.}
\begin{ruledtabular}
\begin{tabular}{lcccc}
Material & $T_{\rm{C}}$  & $\mu_0H(5\%)$  &  $\mu_0H(8\%)$ & $k_{\rm{B}}T_{\rm{C}}/gJ\mu_{\rm{B}}$ (T) \\
         &   (K)    &      (T)       &   (T)          &  (T)                 \\
\colrule
Gd$_5$Si$_4$        & 336 & 5.86  & 11.87 & 71.4    \\
Gd                  & 293 & 5.08  & 10.33 & 62.3    \\
Gd$_3$In            & 213 & 3.62  & 7.45  & 45.3     \\
GdMg                & 119 & 1.76  & 2.94  & 25.3     \\
GdRu$_2$            & 84  & 0.54  & 2.43  & 17.9      \\
GdPd                & 46  & 0.32  & 1.30  & 9.8     \\
\end{tabular}
\end{ruledtabular}
\end{table}

First of all, if we think in very basic terms there is one magnitude that will certainly induce the destruction of the scaling behavior of the magnetic entropy change peak. For some value of applied field, a very close value to the theoretical limit of $\Delta S_{\rm{M}}$ is going to be achieved. For this large enough value of applied field all magnetic spins are going to be aligned along the same direction. Then, if at that point we increase the applied field we are not going to notoriously increase the value of the magnetic entropy change because it is indeed very close to the maximum value. Hence, in this regime of very high applied fields, the value of the exponent $n$ is going to be 0 or very close to it. This is in agreement with the decrease from 0.66 to 0.5 showed in Fig. \ref{fig2} for the cases of GdRu$_2$ and GdPd. For this reason, it is obvious that for some value of the applied field the exponent $n$ at $T_{\rm{C}}$ will begin to decrease from the predicted value of the critical exponents to 0. Now, we want to elucidate which are the values of these fields for which the power laws are not valid any more. However, according to the results showed in Fig. \ref{fig2}, we cannot focus exclusively on the magnetic field because the deviations from the ideal behavior are only apparent for the low Curie temperature materials (when we restrict ourselves to the same values of applied field). Therefore, both magnitudes have to be involved in the worsening of the scaling behavior. In order to extract some quantitative conclusions, we analyze the field dependence of the magnetic entropy change peak for all these materials. In Sec. \ref{meanfieldtheory} we have showed the expression of Eq. (\ref{fielddep}) for the theoretical field dependence of the magnetic entropy change peak near the phase transition, so we can compare this ideal behavior (which is a power law) with the real behavior obtained by Eq. (\ref{brillouin}). In this way, we can observe simultaneously which is the effect of both quantities. On the one hand, we analyze for which applied field breakdown of the power laws appears, on the other hand, by comparing different materials with different working temperatures, we analyze if this applied-field limit has a tendency with respect to the Curie temperature. All these results are collected in Fig. \ref{fig3}, where we have marked, for each material, a couple of values of applied field, labeled by $H(5\%)$ and $H(8\%),$ which represent those values for which the power law prediction of the maximum magnetic entropy change differs only by 5\% and 8\% from the real one, respectively. The 5\% limit has been chosen taking into account the typical error margin of experimental MCE measurements. The 8\% limit represents deviations from the scaling behavior which cannot be ascribed to experimental uncertainty. Although these two limiting values can be considered somewhat arbitrary, especially the upper one, their modification would not alter the conclusions of this work. According to this, for applied field lower than $H(5\%)$ we can apply all power laws in a reliable way. For applied fields between $H(5\%)$ and $H(8\%)$, we could apply them too, but it is possible that we have non-negligible errors. And for applied fields above $H(8\%),$ we should not apply any power law at all because the differences are too large. As we can see in Fig. \ref{fig3}, these reference values of applied field are different for each material and, as expected, these values are clearly related to the Curie temperature. In Table \ref{table2}, we have collected all these values and compared them with an intrinsic field given by $k_{\rm{B}}T_{\rm{C}}/gJ\mu_{\rm{B}}.$ This quantity provides information about the ratio between the magnetic energy and thermal energy at $T_{\rm{C}}.$ According to the data of Table \ref{table2}, the power laws are only valid when the applied field is around one order of magnitude $(\sim8\%)$ smaller than this intrinsic field of the material. To give some useful numbers, for materials with near room temperature transition all scaling approximations are valid even for applied field up to around 10 T. However, for those materials with Curie temperatures below the nitrogen boiling point, this kind of approximations have a limited applicability for applied fields of $\sim1$ T.

We can also analyze the correlation between the scaling behavior and the collapse onto the universal curve of all magnetic entropy change curves for different applied fields. In Sec. \ref{scaling} we have discussed in detail that collapse of the normalized magnetic entropy change curves is direct consequence of the scaling behavior of $\Delta S_{\rm{M}}.$ In Fig. \ref{fig4} we collect the universal curves for the six studied materials. As expected, the dispersion of the universal curve for $\theta<0$ (the ferromagnetic region) remains small only for those materials with large Curie temperature and it becomes large for the rest, where significant deviations from the scaling laws were observed in the exponent $n.$ This kind of deviation from the ideal behavior predicted by scaling laws has been observed before in measurements of low Curie temperature materials. See, for example, the work by Y. Su et al. \cite{su2013} on YbTiO$_3,$ with $T_{\rm{C}} = 42$ K, and related perovskites or the recent paper by L. Li et al. \cite{li2015b}, which deals with HoZn intermetallic compound with $T_{\rm{C}} \sim 72$ K. Our results point out that the presence of additional magnetic phase transition is not the only cause of the lack of universal scaling at low temperatures.

\subsection{Experimental support}\label{experimentsresults}

\begin{figure}[t]
\includegraphics[width=0.5\textwidth]{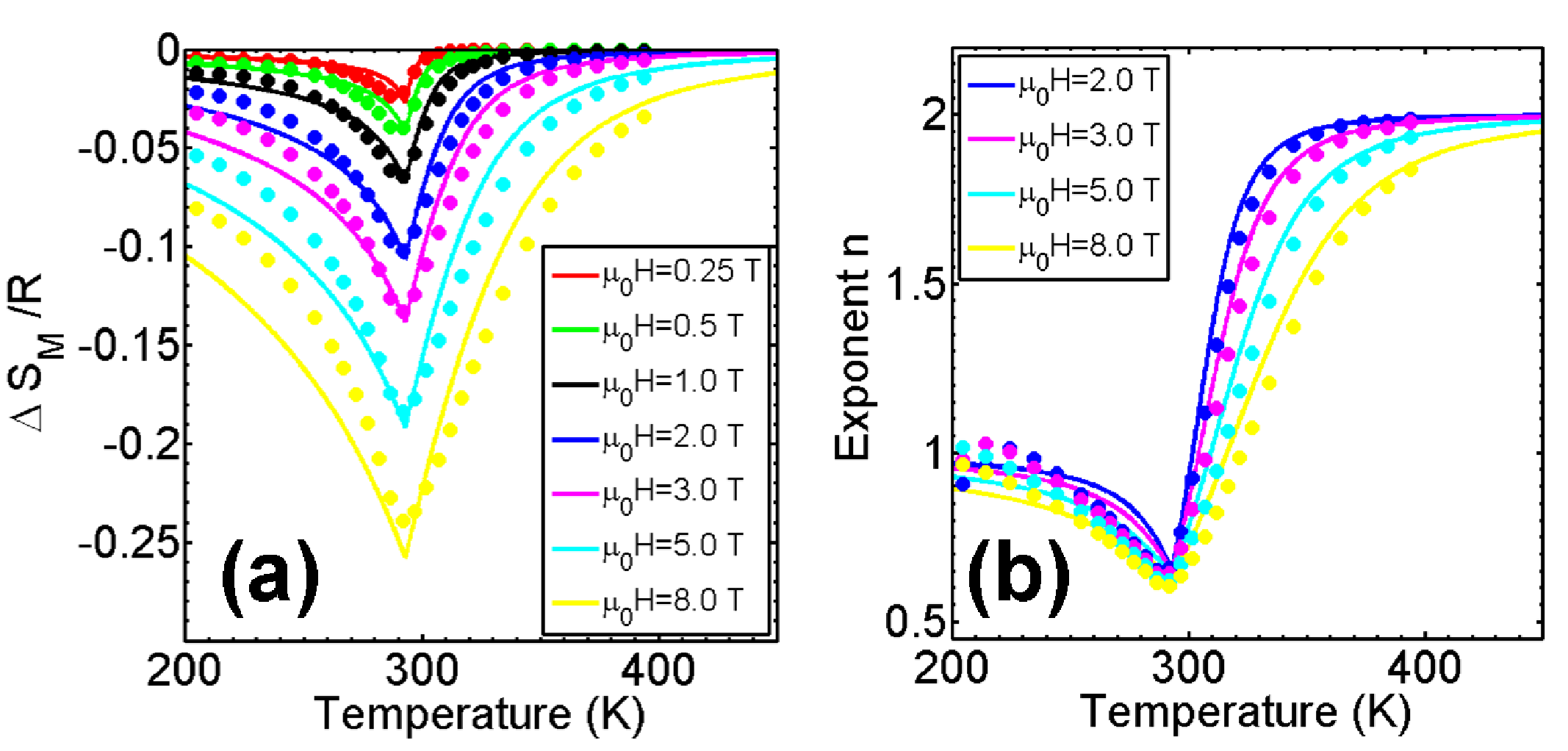}
\caption{\label{fig5} Comparison between the calculated magnetic entropy change curves with the mean-field Brillouin model (solid lines) and the experimental data of gadolinium with applied fields up to 9 T (symbols) (a). The same comparison between the exponent $n$ obtained with Eq. (\ref{exponent}) (b). In both cases experimental data are corrected with a demagnetizing factor of 0.26. This correction is especially relevant in the calculation of the exponent $n.$ Despite of the simplicity of the model, the agreement is quite reasonable.
}
\end{figure}

\begin{figure}[t]
\includegraphics[width=0.5\textwidth]{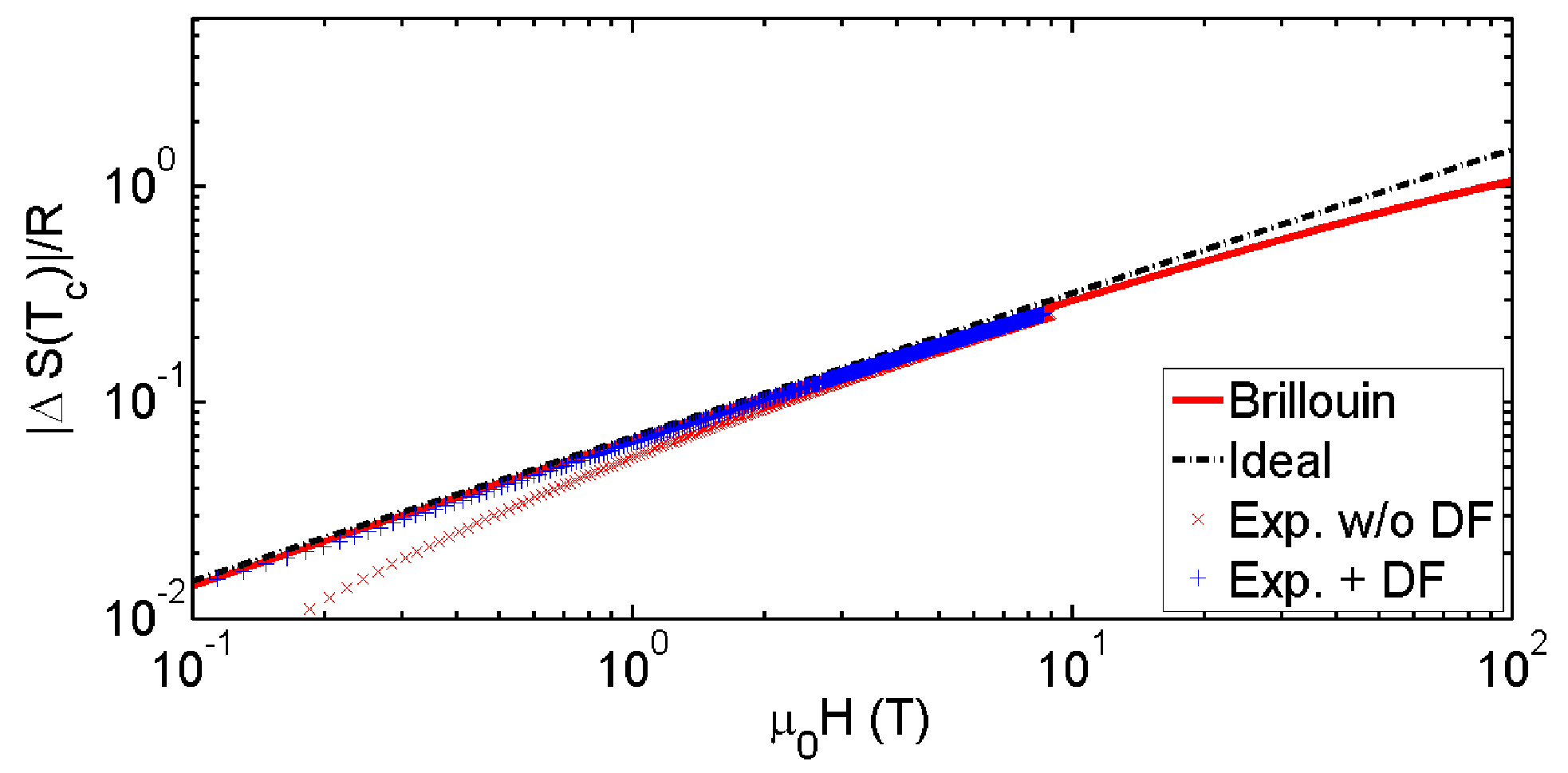}
\caption{\label{fig6} Comparison of the field dependence of the magnetic entropy change peak between the mean-field Brillouin model and the experimental data with and without demagnetization factor (DF). As it was predicted by the model for the case of gadolinium whose Curie temperature is at room temperature the power law for the field dependence is fulfilled up to fields of 10 T.
}
\end{figure}

As it was previously pointed out, with the Brillouin function the magnetothermal response of a simple ferromagnetic material can be reproduced with reasonable accuracy. To show this, Fig. \ref{fig5} shows a comparison between the magnetic entropy change and $n(T)$ curves obtained with the mean-field Brillouin model and experimental data for a flat disk shaped piece of Gd measured in a vibrating sample magnetometer for different applied fields up to 9 T. Despite of the simplicity of the model, it is obvious that the agreement is quite good, at least around the Curie temperature. In both cases experimental data were corrected with a demagnetizing factor of 0.26 in SI untis. Notice that in the case of the experimental data, even if the shape of the sample (a thin plate measured with the field applied in the plane of the sample) has been chosen to minimize the influence of the demagnetizing factor, a contribution of this demagnetizing field is known to alter the MCE of materials change peak \cite{caballeroflores2009}. This effect is especially relevant in the $n(T)$ curves  where $n=1$ limits would not be achieved for $T\ll T_{\rm{C}}$ in the presence of demagnetizing fields \cite{romeromuniz2014} as in this case does. With the aid of the experimental data we can check our proposal of the limits of validity of scaling relations obtained by the mean-field Brillouin model. According to our previous results for gadolinium, whose Curie temperature is 293 K, in the proximity of room temperature, we should expect a good power law behavior for applied fields up to 10 T. In Fig. \ref{fig6} we show the field dependence of the magnetic entropy change peak obtained in the mean-field Brillouin model together with the result of experimental data. Notice how the correction of the demagnetization factor is completely necessary \cite{romeromuniz2014}, otherwise, we would find a relevant discrepancy in the low-field region but completely artificial and unrelated to the theoretical model. The calculated maximum of magnetic entropy change is slightly lower $(<\!\!4\%)$ than the experimental one but they are close enough. Another important point that must be clarified is that critical exponents of real gadolinium ($\beta = 0.39$ and $\gamma = 1.24$) \cite{srinath1999} are significantly different from those of mean-field values.  However, it is possible to compare directly all results because the exponent $n$ in the case of real gadolinium is 0.626, which very close to the 2/3 value of the mean-field theory. Therefore, no relevant difference is expected in the slope of the critical scaling when plotting both sets of data. We see how up to applied fields of 9 T the agreement between calculated data by the mean-field Brillouin model and experimental data is excellent. Consequently, as it is shown experimentally it is possible to apply scaling relations in the maximum entropy change of the MCE in the usual working applied fields for materials with Curie temperature close to room temperature. This fact can be used to estimate the performance of different materials without carrying out experimental measurements at very high applied fields.

\subsection{3D-Ising model}\label{isingresults}

It is important to complete our results using the 3D-Ising model introduced in Sec. \ref{isingtheory} since different results have been often observed in models beyond the mean-field approximation. Regarding the MCE, we can find several examples, for instance, in all magnetic equations of state based on mean-field approximation \cite{weiss1907,bean1962,kuzmin2008} the exponent $n$ for the applied field dependence of the magnetic entropy change at the peak is going to be 2/3, as predicted by mean-field critical exponents. On the contrary, in other models beyond the mean-field approximation, this value is different and it is given in terms of the critical exponents. Other example is the peak position of the magnetic entropy change. While in the mean-field approximations the temperature of the peak coincides with the Curie temperature, in the models beyond the mean-field approximation the peak temperature is located slightly above the Curie temperature and linearly increases with the applied field \cite{franco2009}.

\begin{figure}[b]
\includegraphics[width=0.4\textwidth]{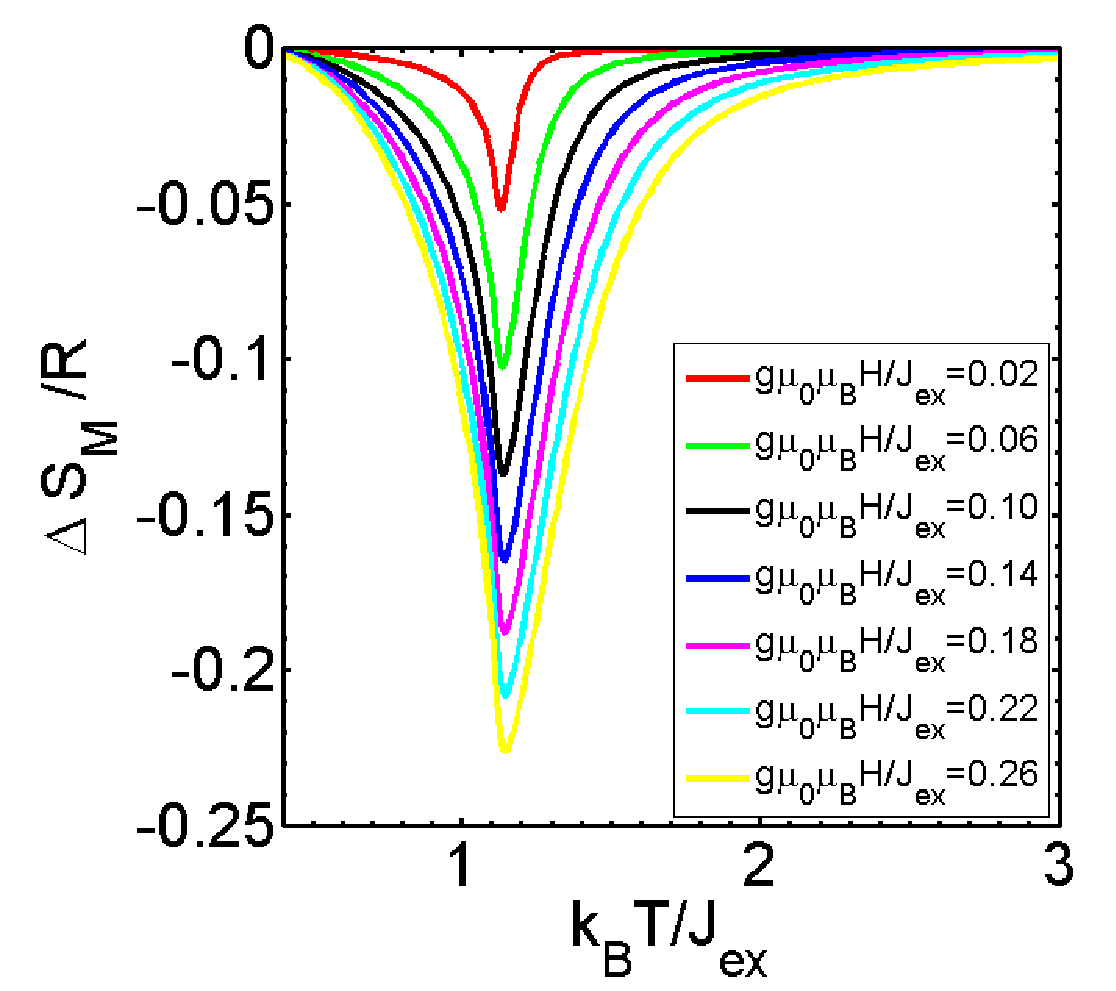}
\caption{\label{fig7} Calculated magnetic entropy change curves with the 3D-Ising model for different applied fields. Notice that in this case we work with dimensionless magnitudes in applied field and temperature.
}
\end{figure}

\begin{figure}[b]
\includegraphics[width=0.4\textwidth]{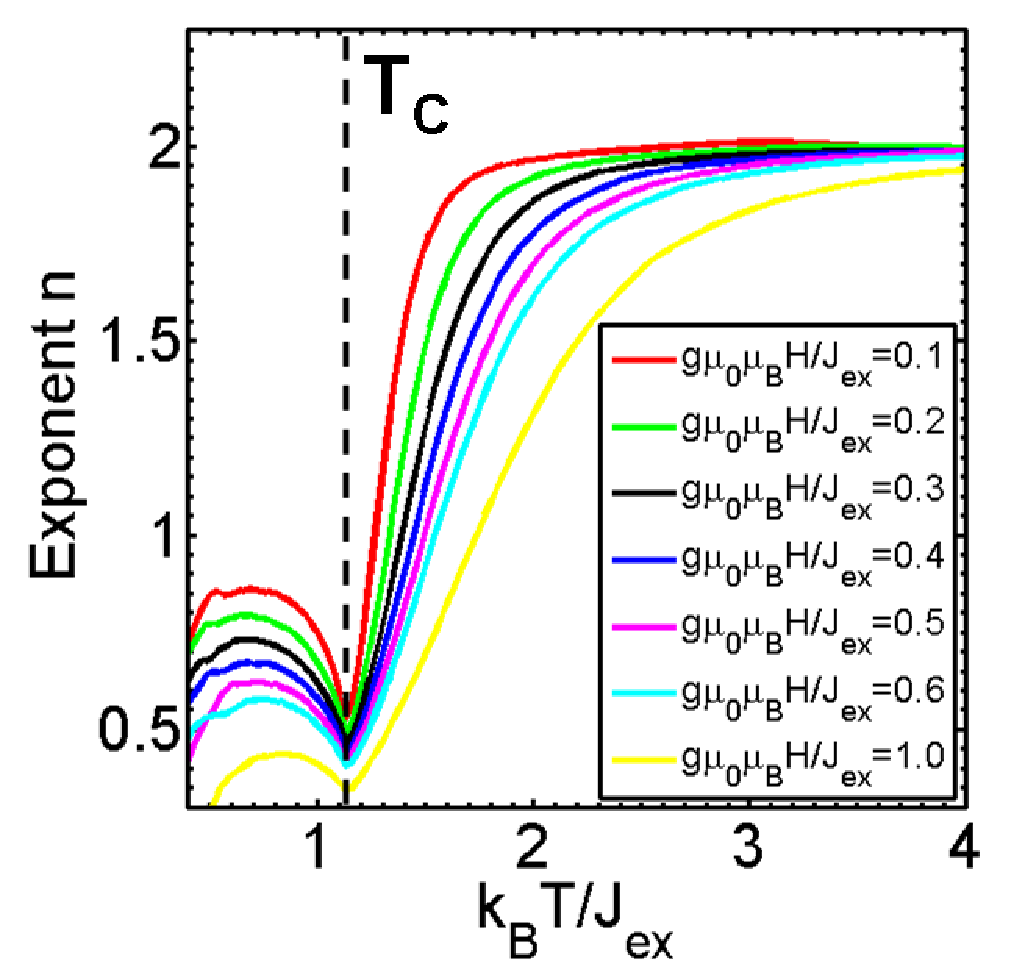}
\caption{\label{fig8} Exponent $n$ with the 3D-Ising model for different applied fields calculated from Eq. (\ref{exponent}). Notice that the behavior is the same as in the mean-field Brillouin model. For high applied field the exponent at $T_{\rm{C}}$ begins to decrease and the limit of $n = 1$ is not satisfied for very low temperatures.
}
\end{figure}

\begin{figure}[t]
\includegraphics[width=0.45\textwidth]{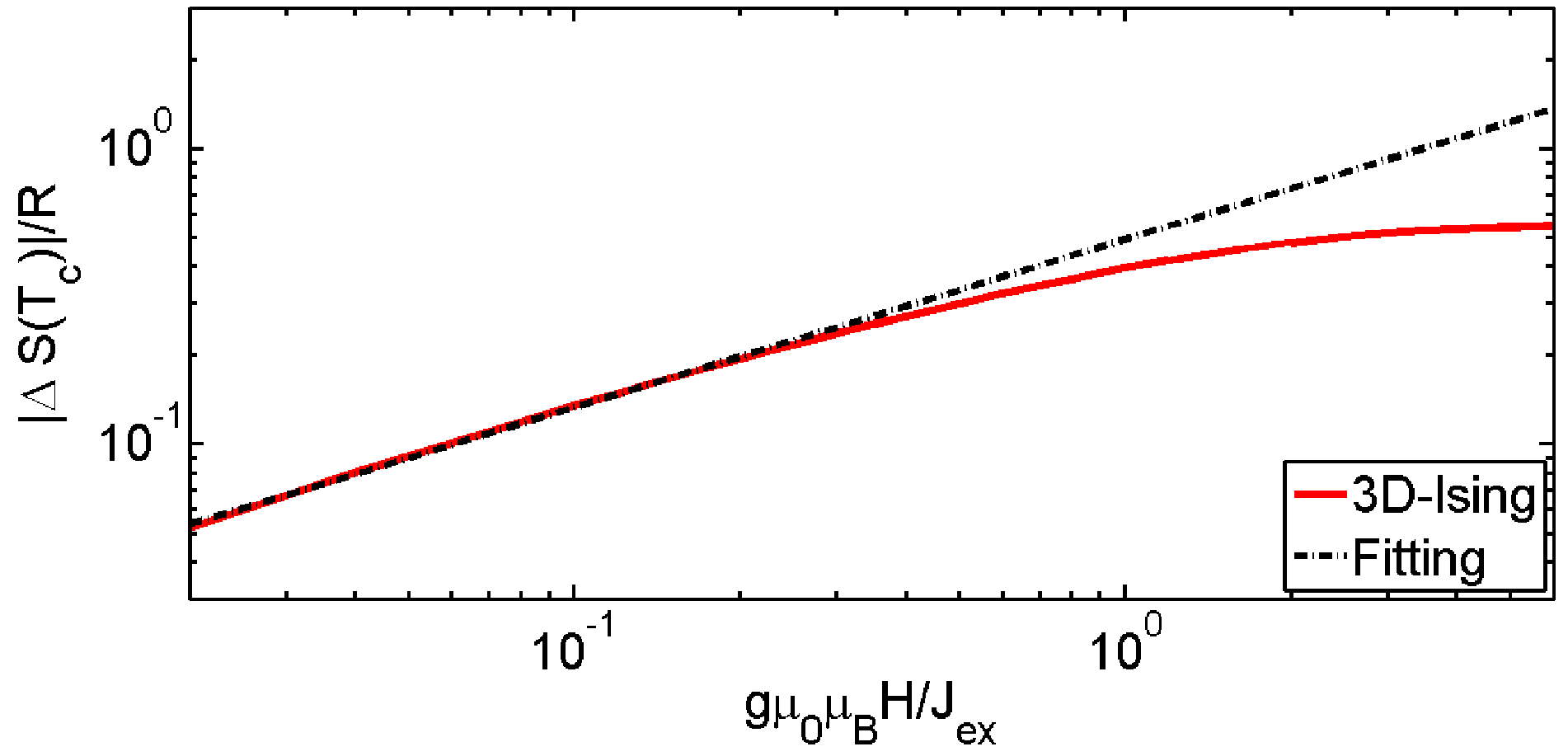}
\caption{\label{fig9} Applied-field dependence of the magnetic entropy change peak in the 3D-Ising model. In this case, an analytical expression for the scaling relation cannot be provided. Instead, we perform a linear fitting in the low applied field region. The exponent obtained in the fitting is very similar to the value predicted by the critical exponents of the model. (see text for details).
}
\end{figure}

\begin{figure}[t]
\includegraphics[width=0.4\textwidth]{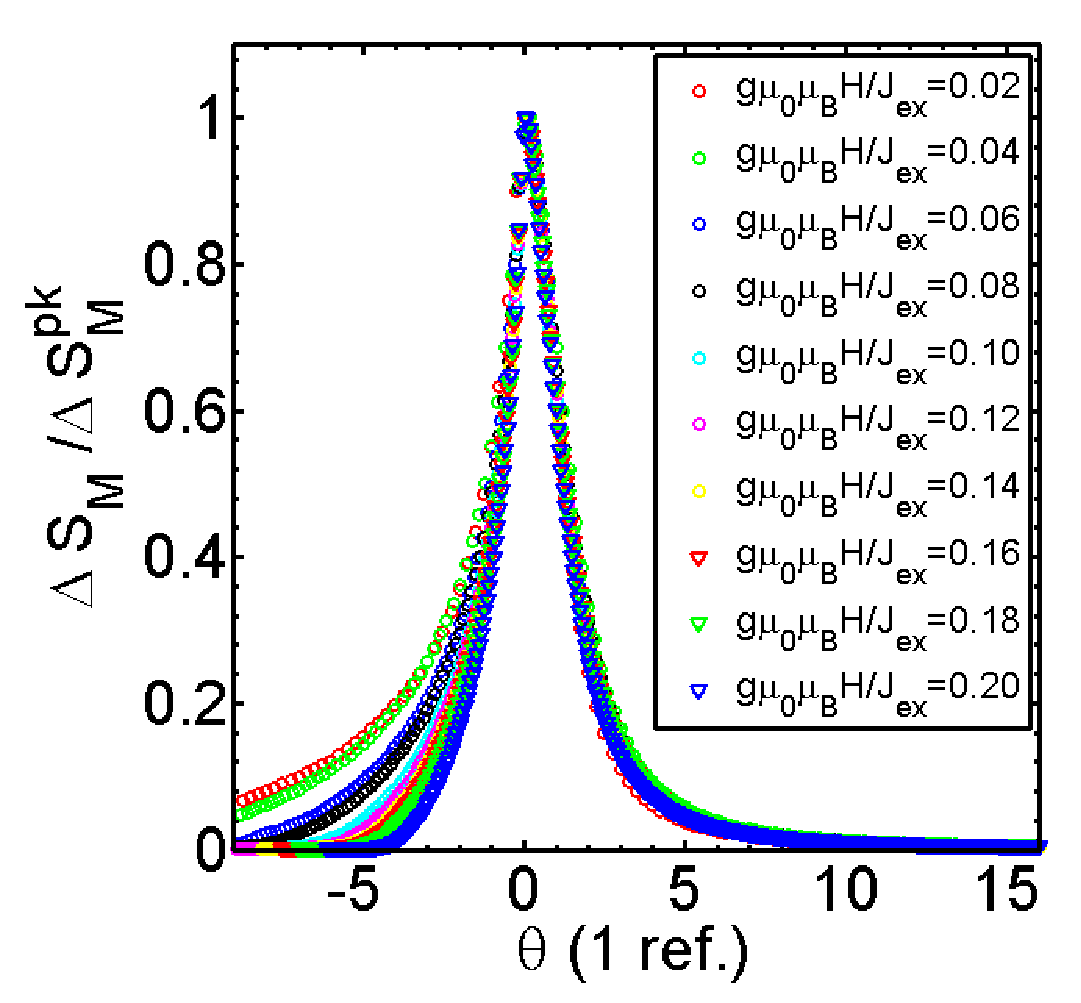}
\caption{\label{fig10} Universal curve of the magnetic entropy change in the 3D-Ising model. The specific value of the Curie temperature of this model produces a quick saturation of the material in the ferromagnetic range, affecting the collapse of the curves as discussed in the text.
}
\end{figure}

The aim of this section is to carry out a similar analysis of the previous sections in the 3D-Ising model where $J_{\rm{ex}}>0$ as explained in Sec. \ref{isingtheory}. The Ising model is a microscopic model and can treat physical properties beyond the mean-field approximation and it belong to a different universality class \cite{hasenbusch2010}. By using the 3D-Ising model, we can confirm that the conclusions extracted in Secs. \ref{meanfieldresults} and \ref{experimentsresults} are completely general. Figure \ref{fig7} shows the temperature dependence of the magnetic entropy change for different applied fields for $L=16,$ which are calculated by the Wang-Landau method. We have chosen values of the applied field so that the usual experimental conditions can be reproduced. In the experimental data of gadolinium presented in Sec. \ref{experimentsresults} the applied field was within the range $1-10$ T. In this paper, $J_{\rm{ex}}$ is the energy unit, so the temperature and applied field are expressed by $k_{\rm{B}}T/J_{\rm{ex}}$ and $g\mu_0\mu_{\rm{B}}H/J_{\rm{ex}}$, respectively. By comparing $k_{\rm{B}}T/J_{\rm{ex}}$ and $g\mu_0\mu_{\rm{B}}H/J_{\rm{ex}},$ we can estimate the value of the applied field, finding that $g\mu_0\mu_{\rm{B}}H/J_{\rm{ex}}>1$ corresponds to over 100 T for the materials whose Curie temperatures are $\sim 100$ K. Thus, in order to reproduce the order of $1-10$ T, it is reasonable to use $g\mu_0\mu_{\rm{B}}H/J_{\rm{ex}}$ in the range of $0.01-0.1.$ The temperature dependence of the exponent $n$ for different applied fields is shown in Fig. \ref{fig8}. The same behaviors observed in the mean-field Brillouin model shown in Fig. \ref{fig2} explained in Sec. \ref{meanfieldresults}, that is, at $T\ll T_{\rm{C}}$ and $T\gg T_{\rm{C}},$ respectively, bounds of $n=1$ and $n=2,$ are obtained in the 3D-Ising model for low applied fields. The different point from the mean-field Brillouin model is that the value of $n$ at $T_{\rm{C}}$ becomes 0.569, which is obtained by the critical exponents of the 3D-Ising model by $n=1+(\beta-1)/(\beta+\gamma)$. However, these behaviors are shifted when the applied field becomes large, because the applied field is so high that we are close to the saturation magnetization regime, as discussed in Sec. \ref{meanfieldresults}. Figure \ref{fig9} shows the magnetic entropy change at the peak $|\Delta S_{\rm{M}}^{\rm{pk}}|,$ plotted in logarithmic scale, as a function of the applied field. In this model, we cannot provide an analytical expression of the magnetic entropy change. However, it is possible to confirm the power-law behavior of this magnitude by performing a linear fitting in the low applied field region. The intrinsic field of this model defined in Sec. \ref{meanfieldresults} is $2k_{\rm{B}}T_{\rm{C}}/J_{\rm{ex}} = 2.254$ (taking into account the value of $J=1/2$ in this model) and the 8\% of this field is 0.18. By performing the linear fitting in this region $g\mu_0\mu_{\rm{B}}H/J_{\rm{ex}}<0.18,$ we obtain $n=0.571$, which is in perfect agreement with the expected value of 0.569 from the critical exponents. This fitting curve is also shown in Fig. \ref{fig9} and notice that expanding the linear fitting to higher values of applied fields the obtained value of exponent $n$ differs dramatically from the expected value.

Finally, in Fig. \ref{fig10}, we plot the magnetic entropy change curves for different applied fields (with a single temperature reference) which are in the range of applicability of the scaling relations (with two extra curves outside of the correct field range). Around the Curie temperature, all magnetic entropy change curves are superposed by using normalized magnetic entropy and normalized temperature but for negative values of $\theta$ not very far away from zero because the Curie temperature is only $k_{\rm{B}}T/J_{\rm{ex}}=1.127$.

\subsection{Other remarks}\label{otherresults}

There are several aspects which might be important when analyzing experimental data and, especially, when comparing with theoretical models. Firstly, we have proved in the previous sections that we have to be cautious if we want to apply the power laws derived for the MCE magnitudes. We have to make sure, first, that we are in the appropriate conditions of temperature and applied field to do this. Additionally, when using theoretical magnetic equations of state we have to be aware that, depending on the model, the scaling relations can be fulfilled in a broader or narrower range. For instance, we have seen that for the Brillouin function the range of applicability is similar to that observed in experiments but it might not be the case for other models. Of course, the power laws should be guaranteed at least for $T \rightarrow T_{\rm{C}}$ if the model is well constructed. However, there are models in which the range of applicability is very narrow, for example in the Kuz'min equation of state \cite{kuzmin2008}. In other cases, like in the Bean-Rodbell model \cite{bean1962}, some additional parameters are included, so they can promote the destruction of these critical phenomena or lead to an artificial narrowing of its validity range. This fact has been interpreted by some researches as a proof of the lack of scaling relations in materials \cite{smith2014}. Finally, we want to pay some attention to those equations of state based directly on scaling relations like the Arrott-Noakes equation \cite{arrott1967} or the Ho-Litster equation \cite{ho1969}. The first one has been widely used among the MCE community with plenty of success. In fact, it has been used to fit experimental data with very high accuracy \cite{franco2008b,franco2011}. The Arrott-Noakes equation of state was derived originally to reproduce the magnetothermal response of pure nickel and can be expressed as:
\begin{equation}
\left(\frac{H}{M}\right)^{1/\gamma} = a(T-T_{\rm{C}}) + bM^{1/\beta},
\end{equation}
where $a$ and $b$ are two fitting parameters. This equation is extremely useful due to its simplicity and the two fitting parameters add a lot of freedom allowing the understanding of the experimental data. Another advantage of this equation is the wide range of applicability, which has been proved to be valid for $t(M/H)^{-1/\beta} \alt 25$ within errors of less than 1\% \cite{compostrini2002}. Moreover, this equation has been used to extract valuable information in the context of MCE such as composite materials \cite{romeromuniz2013} or the demagnetizing factor \cite{romeromuniz2014}. Nevertheless, it is worth noting that this type of equations of state are constructed by applying the critical scaling in the whole range of applied field and temperature, which means that in these models a real saturation in magnetization is never reached. This fact has the clear consequence that the exponent $n$ is completely independent of the applied field because no saturation in magnetization is taken into account.

\begin{figure}[t]
\includegraphics[width=0.5\textwidth]{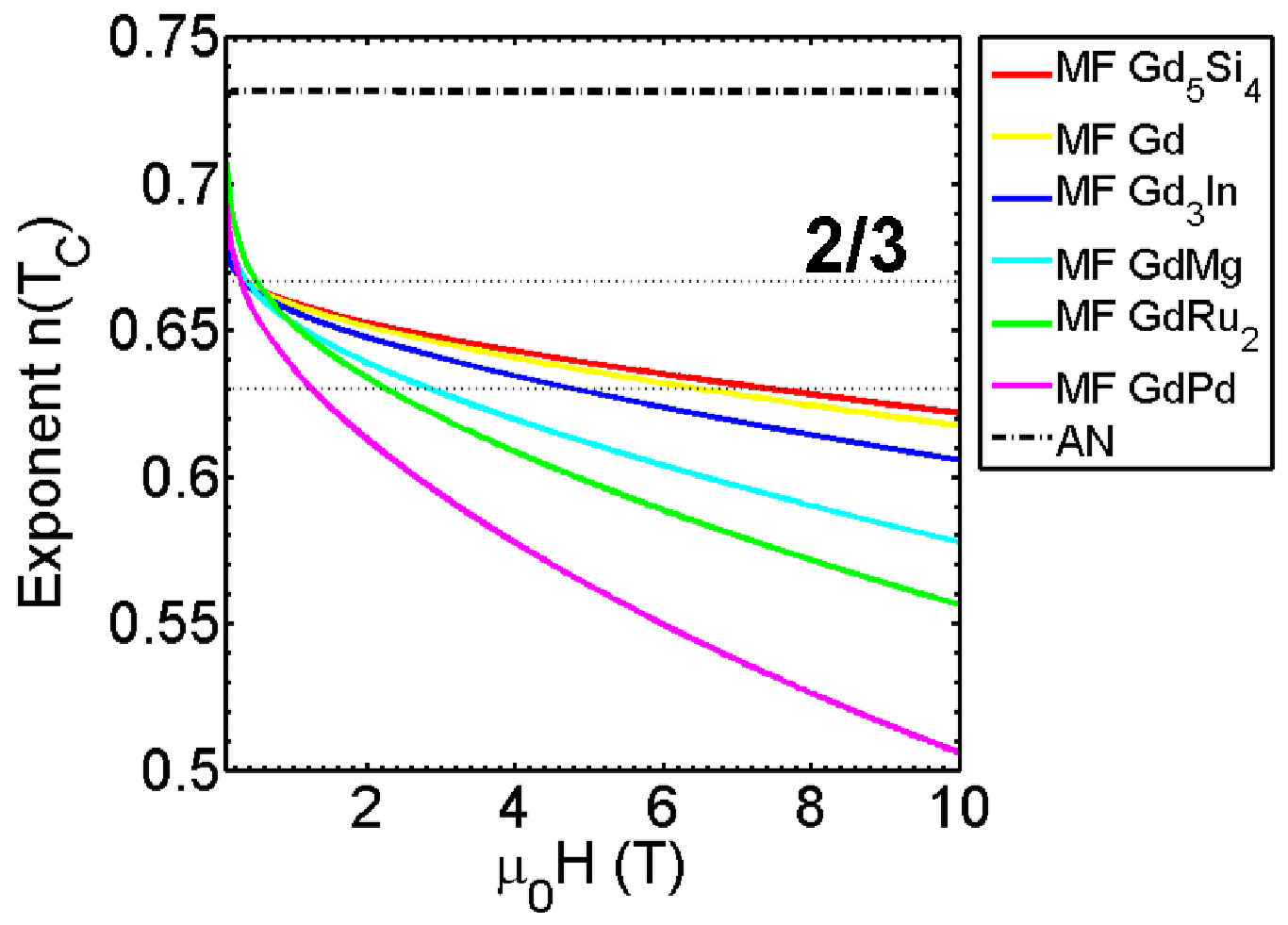}
\caption{\label{fig11} Applied-field dependence of the exponent $n$ evaluated at $T_{\rm{C}}$ for the different materials calculated with the mean-field Brillouin model (MF). Notice the slight dependence with respect to the applied field for those materials with low Curie temperature. For materials with near room Curie temperature the values of $n$ remain close to the predicted value. On the contrary, for scaling models like the Arrott-Noakes (AN) equation there is no field dependence at all. Dashed lines correspond to the ideal value of $2/3$ and to a 5\% lower value which would be acceptable.
}
\end{figure}

As an example we show in Fig. \ref{fig11} the applied field dependence of the exponent $n$ evaluated at $T_{\rm{C}}$ for the mean-field model and for a soft amorphous alloy modeled by the Arrott-Noakes equation \cite{franco2008b} with $n = 0.7313.$ As it is shown in the graph, with the mean-field approximation of the Brillouin function a slight field dependence exists in the exponent $n$ at $T_{\rm{C}}$ especially for materials with low Curie temperatures as we discussed in Sec. \ref{meanfieldresults}. On the contrary for the alloy modeled with the Arrott-Noakes equation of state the exponent $n$ is constant all over the whole range of applied fields. As we pointed out, this incorrect tendency is due to the absence of a real saturation of magnetization. However, if we remember the experimental data of gadolinium in Fig. \ref{fig5}, in the usual experimental applied-field range we are still far from the complete saturation region. For this reason, it is completely acceptable the use of this type of equations in the standard experimental conditions and in the neighborhood of the phase transition.

\section{Conclusions}

In this work we have carried out a detailed study of the field dependence of the magnetic entropy change by means of a mean-field Brillouin model and a 3D-Ising model. We have shown that the field dependence of the magnetic entropy change at the peak obeys a power law according to the critical exponents of each model but not in the whole range of applied field. Therefore, we have delimited which is the range of applicability of this behavior in terms of temperature and applied field. We have proven that, even for temperatures very  close to the Curie temperature for high enough applied fields, the exponent $n$ starts to decrease, reaching zero value for infinitely high fields. This is due to the achievement of the total orientation of the domains in the direction of the applied field, reaching the theoretical maximum in magnetic entropy change, $R\ln(2J+1),$ regardless of any further increase of the applied field. On the other hand for temperatures well below the Curie temperature the scaling behavior is also lost even for moderate values of applied fields. According to these results, the scaling behavior of the MCE is only valid when the energy arising from the magnetic field $\mu_0gJ\mu_{\rm{B}}H$ is much less than the energy contribution of the temperature $k_{\rm{B}}T.$ In practical terms this means that for materials with phase transition near room temperatures these scaling approximations are valid in a broad enough applied field range $(\lesssim10$ T) so the use of the power laws is completely justified and it could be very useful in magnetic measurements. This point was confirmed with experimental data of gadolinium. However, for materials with transition temperature of the order of the boiling point of nitrogen or less, we should be very careful when applying these scaling relations because they will not be valid even for such low applied fields of the order of 1 T.

\begin{acknowledgments}
This work was supported by JSPS KAKENHI Grant Numbers 25420698 (R.T. and S.T.), 15K17720 (S.T.), 15H03699 (S.T.), and by the Spanish MINECO, EU FEDER (Project MAT2013-45165-P) and the PAI of the Regional Government of Andaluc\'{i}a (V.F.). S.T. was also supported by Waseda University Grant for Special Research Projects (Project number: 2015B-514) and C.R.-M. is grateful to FPI-UAM graduate scholarship program and Fundaci\'{o}n Universia for financial support. The computations for the 3D-Ising model were performed on super computers at National Institute for Materials Science, Yukawa
Institute for Theoretical Physics, and Supercomputer Center, Institute for Solid State Physics, the University of Tokyo.
\end{acknowledgments}

\providecommand{\noopsort}[1]{}\providecommand{\singleletter}[1]{#1}%

\end{document}